\documentclass[12pt]{article}
\usepackage{amsmath,latexsym,exscale,amssymb,amsthm}
\newcommand {\al}   {\alpha}       \newcommand {\bt}  {\beta}
       \newcommand {\G }  {\Gamma}
\newcommand {\dl}   {\delta}       \newcommand {\e }  {\epsilon}
        \newcommand {\et}  {\eta}
\newcommand {\lm}   {\lambda}      \newcommand {\m }  {\mu}
\newcommand {\n }   {\nu}          
\newcommand {\s }  {\sigma}

         \newcommand {\om}  {\omega}

\newcommand {\pl}   {\partial}     \newcommand {\nb}  {\nabla}

\newtheorem {Theorem}  {Theorem}

\begin{document}
\title     {New constraints \\
          in dynamical torsion theory}
\author    {M. O. Katanaev
            \thanks{E-mail: katanaev@mi.ras.su}\\ \\
            \sl Steklov Mathematical Institute,\\
            \sl Gubkin St., 8, 117966, Moscow, Russia}
\date      {16 September 1992}
\maketitle
\begin{abstract}
	     The most general Lagrangian for dynamical torsion theory
	     quadratic in curvature and torsion is considered.
	     We impose two simple and physically reasonable constraints
	     on the solutions of the equations of motion: (i) there
	     must be solutions with zero curvature and nontrivial
	     torsion and (ii) there must be solutions with zero torsion
	     and non covariantly constant curvature. The constraints
	     reduce the number of independent
	     coupling constants from ten to five. The resulting theory
	     contains Einstein's general relativity and Weitzenb\"ock's
	     absolute parallelism theory as the two sectors.
\end{abstract}
\section{Introduction}
Dynamical torsion theory (known also as a Poincar\'e gauge theory)
is the simplest geometric generalization of general relativity in
which torsion as well as metric is an independent dynamical
variable \cite{Hehl80}. The most general ten parameter invariant Lagrangian
yielding second order equations of motion contains invariants
quadratic in curvature and torsion. Large number of independent
coupling constants raises a crucial question what choice of the
coupling constants are most acceptable from physical point of view?

There have been different approaches to get restrictions on the
coupling constants in dynamical torsion theory. Unfortunately, there
are weak experimental bounds (see, for example, \cite{Nevill82,Stoege85})
and one has to take into account theoretical considerations. In [4--6]
\nocite{SezNie80,Sezgin81,KuhNit86} propagators for all modes entering
the theory were considered and Lagrangians without ghosts and tachyons
were found. Closely related criterion of positivity of masses and
energy contributions to the canonical Hamiltonian in the linear
approximation was considered in \cite{HayShi80D,MiNaOhTa81}. Another
constraints on the coupling constants can be found by the validity of
Birkhoff theorem [9--11]. \nocite{RamYas79,Nevill80C,RaShNi82}
The restrictions on the coupling constants of the quadratic torsion
terms can be obtained by the requirement of asymptotically Newtonian
behavior of the gravitational field \cite{ChChHsNeYe88}. The initial
value problem and the corresponding restrictions on the coupling
constants following from the validity of Cauchy--Kowalevski theorem
as well as the hyperbolicity conditions were found in
\cite{Dimaki89A,Dimaki89B}. To obtain the above restrictions on the
coupling constants one has to make cumbersome calculations.
Therefore, simpler restrictions are plausible.

In the present paper we impose new simple but very restrictive
constraints on the Lagrangian. Namely, we demand equations of motion
to admit solutions (i) with zero curvature and nontrivial torsion
(absolute parallelism) and (ii) with zero torsion and non covariantly
constant curvature (Einsteinian limit). We prove that only five
parameter Lagrangian satisfies these constraints. The resulting
Lagrangian is the sum of the Hilbert--Einstein Lagrangian for tetrad
field, the three invariants quadratic in curvature, and cosmological
constant. The three quadratic in curvature terms entering the
Lagrangian are those that vanish in the case of zero torsion and
describe massless Lorentz connection. It is very interesting that the
resulting Lagrangian, after further fixing one more constant and
dropping the cosmological constant, coincides with the unique
Lagrangian without ghosts and tachyons found by Kuhfuss and Nitsch
\cite{KuhNit86}.

These constraints are motivated by the following consideration.
Equations of motion in a general dynamical torsion theory have a
very particular form. If one sets torsion equal zero in the equations
of motion then one obtains the Einstein equations for a metric and
additional constraint on the curvature tensor that is absent in
general relativity. The last follows from the equation for the
Lorentz connection and states that the curvature must be covariantly
constant. Thus if one wants the usual general relativity to be
incorporated in dynamical torsion theory then one should try to find
the set of coupling constants that admits solutions of zero torsion
and non covariantly constant curvature.

For zero curvature the equation for the Lorentz connection in a
general case yields zero torsion condition, the theory becoming
trivial. Thus if one wants the
Weitzenb\"ock absolute parallelism theory \cite{HayShi79,MulNit85} to
be also incorporated in dynamical torsion theory then one should find
the coupling constants that admit solutions of zero curvature and
nontrivial torsion. It is quite surprising that the set of coupling
constants admitting general relativity and absolute parallelism theory
as the two limiting cases of dynamical torsion theory exists.

The constraints used to reduce the number of coupling constants have
also direct physical interpretation in solids with defects.
It is known that curvature and torsion are correspondingly the surface
densities of Frank and Burgers vectors for continuously distributed
dislocations and disclinations [17--20].
\nocite{Kroner81,KadEde83,McHeMi90,KatVol92}
Thus the two imposed constraints are nothing more than existence
of the space-times with only dislocations or only disclinations. Since
dislocations and disclinations in media are occurred independently, one
may say that the constraints have experimental background.

The paper is organized as follows. In Section~\ref{sLagra} the
equations of motion are derived for a general type action.
In Section~\ref{sconst} we solve the constraints and derive the
five parameter Lagrangian. The Conclusion contains brief
discussion of the resulting model.
\section{General type action}                           \label{sLagra}
Let ${\mathbb R}^4$ be a four-dimensional manifold (space-time) with
coordinates $x^\m,~\m=0,1,2,3$. We assume that the space-time is a
Riemann--Cartan manifold equipped with metric $g_{\m\n}(x)=g_{\n\m}(x)$
and torsion $T_{\m\n}{}^\rho(x)=-T_{\n\m}{}^\rho(x)$ \cite{KobNom63}.
Equivalent realization of Riemann--Cartan geometry can be given in
terms of the tetrad $e_\m{}^a,~a=0,1,2,3$ and the Lorentz connection
$\om_\m{}^{ab}=-\om_\m{}^{ba}$ \cite{HeHeKeNe76}. Torsion and
curvature have the following form in terms of these variables
\begin{eqnarray*}
	T_{\m\n}{}^a&=&\pl_\m e_\n{}^a-\om_\m{}^{ab}e_{\n b}
			  -(\m\leftrightarrow\n),
\\
	R_{\m\n}{}^{ab}&=&\pl_\m\om_\n{}^{ab}-\om_\m{}^{ac}\om_{\n c}{}^b
                       -(\m\leftrightarrow\n).
\end{eqnarray*}

Here and in what follows transition between "curved" Greek and
"flat" Latin indices is performed using the tetrad. Greek indices are
raised and lowered by the metric $g_{\m\n}$, while Latin indices are
raised and lowered by the Lorentz metric $\et_{ab}=$${\rm diag}(+---)$.

The most general invariant action yielding second order equations of
motion for tetrad and Lorentz connection contains ten arbitrary
parameters
\begin{equation}                                        \label{eactio}
	I=\int d^4xeL,~~~~e={\rm det}e_\m{}^a,
\end{equation}
where
\begin{eqnarray}                                        \label{eLagra}
     L&=&\kappa R-\frac14T_{abc}\left(\bt_1T^{abc}+\bt_2T^{cab}
	   +\bt_3\et^{ac}T^b\right)
\\                                                           \nonumber
	&&-\frac14R_{abcd}\left(\al_1R^{abcd}+\al_2R^{cdab}+\al_3R^{acbd}
	  +\al_4\et^{bd}R^{ac}+\al_5\et^{bd}R^{ca}\right)+\lm.
\end{eqnarray}
Ricci tensor, scalar curvature, and the trace of torsion are defined
as follows $R_{ac}=R_{abc}{}^b,~R=R_a{}^a,~T_b=T_{ab}{}^a$. Lagrangian
(\ref{eLagra}) contains all independent invariants constructed from
$T_{abc},~R_{abcd}$ and containing no more than two partial derivatives
\cite{Christ80}.

Let us note two identities which exclude two possible invariants from
the Lagrangian. One has no need to add the Hilbert--Einstein
Lagrangian $\widetilde R(e)$ constructed only from tetrad (scalar
curvature of zero torsion) because of the identity
\begin{equation}                                        \label{ecurid}
	R+\frac14T_{abc}T^{abc}-\frac12T_{abc}T^{cab}-T_aT^a
	-\frac2e\pl_\m(eT^\m)=\widetilde R(e).
\end{equation}
The second identity is the Gauss--Bonnet formula which excludes scalar
curvature squared term
\begin{equation}                                        \label{eGauBo}
	-R_{abcd}R^{cdab}+4R_{ab}R^{ba}-R^2=\frac1e\pl_\m(\dots).
\end{equation}

In the next section we will also use the Bianchi
identities\footnote{In the journal version of the paper the covariant
derivative in the Bianchi identities was understood with the Lorentz
connection acting on Latin indices and the Christoffel symbol acting on
Greek ones. In that case terms with torsion in the right hand sides are
abcent.}
\begin{align}                                        \label{eBian1}
  \nb_\m R_{\n\rho}{}^{ab}+\nb_\n R_{\rho\m}{}^{ab}
  +\nb_\rho R_{\m\n}{}^{ab}=&T_{\mu\nu}{}^\s R_{\rho\s}{}^{ab}
  +T_{\nu\rho}{}^\s R_{\mu\s}{}^{ab}+T_{\rho\mu}{}^\s R_{\nu\s}{}^{ab},
\\                                                      \nonumber
  \nb_\m T_{\n\rho}{}^a+\nb_\n T_{\rho\m}{}^a+\nb_\rho T_{\m\n}{}^a
  =&T_{\mu\nu}{}^\s T_{\rho\s}{}^a
  +T_{\nu\rho}{}^\s T_{\mu\s}{}^a+T_{\rho\mu}{}^\s T_{\nu\s}{}^a
\\                                                      \label{eBian2}
  &+R_{\m\n\rho}{}^a+R_{\n\rho\m}{}^a+R_{\rho\m\n}{}^a,
\end{align}
where $\nb$ denotes the covariant derivative with Lorentz connection
for Latin indices and the corresponding metrical connection
$\G_{\m\n}{}^\rho$ for Greek indices. The latter is defined by equation
\begin{equation}                                        \label{emecon}
	\nb_\m e_\n{}^a=\pl_\m e_\n{}^a-\om_\m{}^a{}_be_\n{}^b
                    -\G_{\m\n}{}^\rho e_\rho{}^a=0.
\end{equation}

Equations of motion for action (\ref{eactio}) have the form
\begin{eqnarray}                                             \nonumber
     \frac1e\frac{\dl I}{\dl e_\m{}^a}&=&\kappa\left(Re^\m{}_a-2R_a{}^\m\right)
     +\bt_1\left(\widetilde\nb_\rho T^{\rho\m}{}_a-\frac14T_{bcd}T^{bcd}e^\m{}_a
	+T^{\m bc}T_{abc}\right)
\\                                                           \nonumber
      &&+\bt_2\left[-\frac12\widetilde\nb_\n(T_a{}^{\m\n}-T_a{}^{\n\m})
	-\frac14T_{bcd}T^{dbc}e^\m{}_a-\frac12T^{b\m c}T_{cab}
	+\frac12T^{bc\m}T_{cab}\right]
\\                                                           \nonumber
      &&+\bt_3\left[-\frac12\widetilde\nb_\n(T^\n e^\m{}_a-T^\m e^\n{}_a)
	-\frac14T_bT^be^\m{}_a+\frac12T^\m T_a+\frac12T^bT_{ab}{}^\m\right]
\\                                                           \nonumber
      &&+\al_1\left(-\frac14R_{bcde}R^{bcde}e^\m{}_a+R^{\m bcd}R_{abcd}\right)
\\                                                      \label{eeqtet}
      &&+\al_2\left(-\frac14R_{bcde}R^{debc}e^\m{}_a+R^{cd\m b}R_{abcd}\right)
\\                                                           \nonumber
      &&+\al_3\left(-\frac14R_{bcde}R^{bdce}e^\m{}_a+\frac12R_{abcd}R^{\m cbd}
	-\frac12R_{abcd}R^{bc\m d}\right)
\\                                                           \nonumber
      &&+\al_4\left(-\frac14R_{bc}R^{bc}e^\m{}_a+\frac12R^{bc}R_{bac}{}^\m
	+\frac12R_{ab}R^{\m b}\right)
\\                                                           \nonumber
      &&+\al_5\left(-\frac14R_{bc}R^{cb}e^\m{}_a+\frac12R^{cb}R_{bac}{}^\m
	+\frac12R_{ab}R^{b\m}\right)+\lm e^\m{}_a=0,
\end{eqnarray}
\begin{eqnarray}                                             \nonumber
	\frac1e\frac{\dl I}{\dl\om_\m{}^{ab}}&=&
     \kappa\left(T_ae^\m{}_b+\frac12T_{ab}{}^\m\right)
	+\bt_1\frac12T^\m{}_{ba}
	+\bt_2\frac14\left(T_a{}^\m{}_b-T_{ab}{}^\m\right)
	+\bt_3\frac14T_be^\m{}_a
\\                                                      \label{eeqcon}
	&&+\al_1\frac12\widetilde\nb_\n R^{\n\m}{}_{ab}
	+\al_2\frac12\widetilde\nb_\n R_{ab}{}^{\n\m}
	+\al_3\frac14\widetilde\nb_\n\left(R^\n{}_a{}^\m{}_b-
	R^\m{}_a{}^\n{}_b\right)
\\                                                           \nonumber
	&&+\al_4\frac14\widetilde\nb_\n\left(R^\n{}_ae^\m{}_b
	-R^\m{}_ae^\n{}_b\right)
	+\al_5\frac14\widetilde\nb_\n\left(R_a{}^\n e^\m{}_b
	-R_a{}^\m e^\n{}_b\right)
\\                                                           \nonumber
	&&-(a\leftrightarrow b)=0.
\end{eqnarray}
Here the tilde sign over the covariant derivative means that it acts
with Christoffell's symbols (metrical connection of zero torsion)
$\widetilde\G_{\m\n}{}^\rho$ on Greek indices and with Lorentz
connection $\om_\m{}^{ab}$ on Latin ones. For example,
$$
     \widetilde\nb_\rho T^{\n\m}{}_a=\pl_\rho T^{\n\m}{}_a
     +\widetilde\G_{\rho\s}{}^\n T^{\s\m}{}_a
     +\widetilde\G_{\rho\s}{}^\m T^{\n\s}{}_a-\om_{\rho a}{}^bT^{\n\m}{}_b.
$$
The difference between Greek and Latin indices arises after
integration by parts because of the identity
$\pl_\m e=e\widetilde\G_{\n\m}{}^\n$.

To analyse the equations of motion we decompose torsion and
curvature into irreducible components. Torsion tensor has three
irreducible components
$$
	T_{abc}=t_{abc}+\e_{abcd}T^{*d}+\frac13(\et_{ac}T_b-\et_{bc}T_a),
$$
where we have extracted totally antisymmetric part by the use of
totally antisymmetric tensor $\e_{abcd},~\e_{0123}=1$, and the trace
\begin{eqnarray*}
	T^{*d}&=&-\frac16T_{abc}\e^{abcd},~~~~~~T_b=T_{ab}{}^a,
\\
	t_{abc}&=&-t_{bca},~~~~~~~~~~~~~~t_{ab}{}^a=0,~~~~~~~~~~
	t_{abc}+t_{bca}+t_{cab}=0.
\end{eqnarray*}

Curvature tensor has six irreducible components
\begin{eqnarray*}
	R_{abcd}&=&C^S_{abcd}+C^A_{abcd}+\frac12\left(R^S_{ac}\et_{bd}
	-R^S_{ad}\et_{bc}-R^S_{bc}\et_{ad}+R^S_{bd}\et_{ac}\right)
\\
	&&+\frac12\left(R^A_{ac}\et_{bd}-R^A_{ad}\et_{bc}-R^A_{bc}\et_{ad}
	+R^A_{bd}\et_{ac}\right)
	+\frac1{12}R\left(\et_{ac}\et_{bd}-\et_{ad}\et_{bc}\right)+\e_{abcd}R^*,
\end{eqnarray*}
where we have extracted totally antisymmetric part $R^*$ and all traces
\begin{eqnarray*}
	R^*&=&-\frac1{24}R_{abcd}\e^{abcd},
\\
	R^S_{ab}&=&\frac12\left(R_{ab}+R_{ba}\right)-\frac14R\et_{ab},~~~~
	R^A_{ab}=\frac12\left(R_{ab}-R_{ba}\right),
\\
	C_{abcd}&=&-C_{bacd}=-C_{abdc},~~~~~~~~~~~~C_{abc}{}^b=0,~~~~~~~~~~
	C_{abcd}+C_{acdb}+C_{adbc}=0,
\\
	C^S_{abcd}&=&\frac12\left(C_{abcd}+C_{cdab}\right),~~~~~~~~~~~~
	C^A_{abcd}=\frac12\left(C_{abcd}-C_{cdab}\right).
\end{eqnarray*}
Irreducible component $C^A_{abcd}$ can be parametrized by a symmetric
traceless pseudotensor
$$
	C^A_{abcd}=-\frac18(\e_{abce}D_d{}^e-\e_{abde}D_c{}^e
	           -\e_{cdae}D_b{}^e+\e_{cdbe}D_a{}^e),
$$
where
$$
	D_{de}=D_{ed}=C^A_{abcd}\e^{abc}{}_e,~~~~~~D_d{}^d=0.
$$
The symmetry of $D_{de}$ can be easily checked by multiplying its
definition by totally antisymmetric tensor.

Note that for zero torsion three irreducible components of curvature
equal zero
$$
	C^A_{abcd}=0,~~~~R^A_{ab}=0,~~~~R^*=0,
$$
and $C^S_{abcd}$ is called the Weyl tensor. In this case the
curvature tensor has additional symmetry
\begin{equation}                                        \label{ecusym}
	R_{abcd}=R_{cdab}.
\end{equation}
\section{Solution of the constraints}                   \label{sconst}
In the present section we impose two constraints on the space of
solutions of equations of motion which allow one to determine five of
the ten parameters entering the Lagrangian (\ref{eLagra}).
The first constraint states that there must be solutions with zero
curvature and nontrivial torsion. The following theorem yields the
restrictions on coupling constants.
\begin{Theorem}                                         \label{ttorsi}
Equation of motion (\ref{eeqcon}) has solution $R_{abcd}=0,
~t_{abc}\ne0,~T^{*a}\ne0,~T_a\ne0$ if and only if the coupling
constants obey
\begin{equation}                                        \label{ebetak}
     \bt_1=-\kappa,~~~~\bt_2=2\kappa,~~~~\bt_3=4\kappa,
\end{equation}
$\al_1,\dots,\al_5$ being arbitrary.
\end{Theorem}
{\bf Proof.} 
Extracting the trace and totally antisymmetric part of equation
(\ref{eeqcon}) for $R_{abcd}=0$ one obtains the equations
\begin{eqnarray}                                             \nonumber
     (8\kappa-2\bt_1+\bt_2-3\bt_3)T_a=0,
\\                                                      \label{etorsi}
     (\kappa-\bt_1-\bt_2)T^{*a}=0,
\\                                                           \nonumber
     (4\kappa+2\bt_1-\bt_2)t_{abc}=0.
\end{eqnarray}
Because all irreducible components of torsion are not equal to zero
one can easily check that Eqs.(\ref{ebetak}) are the unique solution
of Eqs.(\ref{etorsi}).
$\Box$ 

Under the restrictions on the coupling constants (\ref{ebetak}) the
first four terms in the Lagrangian (\ref{eLagra}) yield the
Hilbert--Einstein Lagrangian due to the identity (\ref{ecurid}).
For this choice of the coupling constants and zero curvature,
equation of motion for the Lorentz connection is identically
satisfied  while equation of motion for the tetrad (\ref{eeqtet})
reduces to the Einstein equation
\begin{equation}                                        \label{eEinst}
    \widetilde R_{\m\n}-\frac12\widetilde Rg_{\m\n}-\frac\lm{2\kappa}g_{\m\n}=0.
\end{equation}
Note that the tetrad field satisfying this equation defines nontrivial
torsion on the manifold while the curvature remains zero. This model
is known as Weitzenb\"ock gravity theory with absolute parallelism (or
teleparallelism theory) \cite{HayShi79,MulNit85}.

The second constraint on the space of solutions is that there must be
solutions with zero torsion and non covariantly constant curvature.
Corresponding restrictions on the coupling constants are given by
the following theorem.
\begin{Theorem}                                         \label{tcurva}
Equation of motion (\ref{eeqcon}) has solution
$T_{abc}=0,~\nb_\m R\ne0,~\nb_\m C_{abc}{}^\m\ne0$
if and only if the coupling constants obey
\begin{equation}                                        \label{egamma}
	\al_3=-2(\al_1+\al_2),~~~~~~\al_5=-\al_4.
\end{equation}
$\kappa$ and $\bt_1,~\bt_2,~\bt_3$ being arbitrary.
\end{Theorem}
{\bf Proof.} 
In the case of zero torsion there are some simplifications. Now
quantities with and without tilde sign coincide, and it will be dropped.
Due to the definition (\ref{emecon})
one can easily transform Greek indices into Latin ones inside the
covariant derivatives. The one more simplification arises from the
additional symmetry of the curvature tensor \ref{ecusym}.

Using the Bianchi identity (\ref{eBian2}) equation for the
Lorentz connection (\ref{eeqcon}) is transformed,
\begin{equation}                                        \label{eeqcot}
	\nb^c\left[\left(\al_1+\al_2+\frac12\al_3\right)R_{abcd}
	+\frac14\left(\al_4+\al_5\right)\left(R_{ac}\et_{bd}-R_{ad}\et_{bc}
	-R_{bc}\et_{ad}+R_{bd}\et_{ac}\right)\right]=0
\end{equation}
where $\nb^c=e^{\m c}\nb_\m$. Taking the trace of this
equation with $\et^{bd}$ and using the contracted Bianchi identity
$\nb_\m R=2\nb_\n R_\m{}^\n$ one obtains
\begin{equation}                                        \label{econtr}
	\frac12\left(\al_1+\al_2+\frac12\al_3+\al_4+\al_5\right)\nb_aR=0.
\end{equation}

Using the Bianchi identity (\ref{eBian1}) in the form
$$
	\nb^cC_{abcd}=\frac12(\nb_aR_{bd}-\nb_bR_{ad})
	-\frac1{12}(\nb_aR\et_{bd}-\nb_bR\et_{ad})
$$
the traceless part of (\ref{eeqcot}) can be rewritten as
\begin{equation}                                        \label{econtl}
	2\left(\al_1+\al_2+\frac12\al_3+\frac14\al_4+\frac14\al_5\right)
	\nb^cC_{abcd}=0.
\end{equation}

For $\nb_aR\ne0$ and $~\nb^cC_{abcd}\ne0$  the relations between coupling
constants (\ref{egamma}) are the unique solution of (\ref{econtr})
and (\ref{econtl}).
$\Box$ 

Note that due to the Bianchi identity the Ricci tensor under the
conditions of Theorem~\ref{tcurva} cannot be covariantly constant
$\nb_\n R_\m{}^\n\ne0$. Thus all irreducible components of the
curvature are nonconstant.

Like in the previous case equation of motion for tetrad field under
the restrictions (\ref{ebetak}) and (\ref{egamma}) and zero torsion
condition reduces to the Einstein equation (\ref{eEinst}). But now
tetrad field will define nontrivial curvature, torsion of the manifold
being zero. This is the Einsteinian limit of the model.

As was already mentioned the conditions of the proved theorems have
direct physical interpretation in the physics of solids with
continuously distributed defects. Theorem~\ref{ttorsi} yields the
necessary condition for the existence of a space-time with only
dislocations, while Theorem~\ref{tcurva} yields the necessary
conditions for the existence of a space-time with only disclinations.
In fact, these conditions are sufficient because in both cases
equation of motion for the tetrad field (\ref{eeqtet}) coincides with
the Einstein equation (\ref{eEinst}) which is known to have nontrivial
solutions.

So the two conditions on the space of solution exclude five of the ten
free parameters, and the resulting Lagrangian has the form
\begin{equation}                                        \label{eLagrf}
     \frac1eL=\kappa\widetilde R(e)+c_1C^A_{abcd}C^{Aabcd}
	+c_2R^A_{ab}R^{Aab}+c_3R^{*2}+\lm
\end{equation}
where $c_1=-\frac14(\al_1-\al_2),~c_2=-\frac12(\al_1-\al_2+\al_4),~
c_3=18(\al_1+\al_2)$.
\section{Conclusion}
So we have proved that two simple and physically reasonable
constraints on the space of solutions of dynamical torsion theory
define five parameters Lagrangian. The resulting
theory contains the Hilbert--Einstein Lagrangian, three specific
square curvature terms (which equal zero in the case of zero torsion)
and cosmological constant. The space of solutions of the equations
of motion contains at least two sectors. The first is the celebrated
Einstein's general relativity and the second is the less known
Weitzenb\"ock's absolute parallelism theory. Both sectors are
described by the same dynamical variable that is the tetrad field,
the Lorentz connection being excluded from the theory.
In the Einsteinian limit the Lorentz connection is the known function
of the tetrad, while in absolute parallelism theory Lorentz connection
is a "pure gauge" and can be set equal to zero. In both cases the
dynamics is governed by the Einstein equation for the tetrad which
defines curvature in the Einsteinian limit and torsion in the
absolute parallelism theory.

The Lagrangian \ref{eLagrf} is highly degenerate. Besides the usual
graviton it describes the massless Lorentz connection modes
\cite{KuhNit86}. From Hamiltonian point of view it contains primary
if-constraints that may define the generators of extra gauge symmetry
\cite{BlaVas87A}. The nonpropagating modes of the theory break 2 of
the 10 Cauchy--Kowalevski conditions which are sufficient conditions
for propagating of all modes in the theory \cite{Dimaki89A}. The
hyperbolicity conditions \cite{Dimaki89A,Dimaki89B} (also obtained
for all modes) in our case lead to further restrictions $\kappa=c_1=c_3=0$,
$c_2$ being arbitrary. Because of the large number of nonpropagating
modes and the possible existence of extra gauge symmetry the
Lagrangian \ref{eLagrf} needs a separate consideration. Nevertheless,
by costruction the space of solutions of the equations of motion
surely contains at least two well defined sectors both goverened by
the Einstein equation which is hyperbolic and possesses well posed
Cauchy problem \cite{Choque87}.

Linear approximation for Lagrangian (\ref{eLagrf}) was elucidated in
\cite{KuhNit86}. It is interesting that for $c_2=-2c_1$ and $\lm=0$
one obtains the unique Lagrangian without ghosts and tachyons found
by Kuhfuss and Nitsch \cite{KuhNit86}. Another attractive feature is
that the Lagrangian \ref{eLagrf} is one of the particular Lagrangians
yielding the asymptotically Newtonian behavior of the gravitational
field found in Ref.\cite{ChChHsNeYe88}. The supersymmetric extension
of this Lagrangian is discussed in \cite{Castro89}.

The constraints discussed in the present paper were also considered
in lower number of dimensions. In three dimensional case, which is
relevant to a static media with dislocations and disclinations, one
obtains the sensible two-parameter Lagrangian \cite{KatVol92}
which has physically reasonable solutions. Two-dimen\-si\-onal gravity
with torsion \cite{KatVol86,KatVol90}, which recently was proved to
be completely integrable [22--24] \nocite{Katana90,Katana91,KumSch92A},
and renormalizable \cite{KumSch92B} admits solutions with zero torsion
(only disclinations) but not admit solutions with constant curvature
(only dislocations). Perhaps, the latter fact is connected with the
absence of everywhere smooth vector field on a sphere, singularity of
the vector field (director) representing disclination.

The author would like to thank I. V. Volovich for careful reading the
manuscript and enlightening discussions.

\end{document}